# Low Emittance Growth in a LEBT with Un-Neutralized Section*

L. Prost[#], J.-P. Carneiro, A. Shemyakin
Fermi National Accelerator Laboratory, Batavia, IL 60510, USA

In a Low Energy Beam Transport line (LEBT), the emittance growth due to the beam's space charge is typically suppressed by way of neutralization from either electrons or ions, which originate from ionization of the background gas. In cases where the beam is chopped, the neutralization pattern usually changes throughout the beginning of the pulse, causing the Twiss parameters to differ significantly from their steady state values, which, in turn, may result in beam losses downstream. For a modest beam perveance, there is an alternative solution, in which the beam is kept un-neutralized in the portion of the LEBT that contains the chopper. The emittance can be nearly preserved if the transition to the un-neutralized section occurs where the beam exhibits low transverse tails. This report introduces the rationale for the proposed scheme and formulates the physical arguments for it as well as its limitations. An experimental realization of the scheme was carried out at Fermilab's PIP2IT where low beam emittance dilution was demonstrated for a 5 mA, 30 keV H- beam.

## I INTRODUCTION

A Low Energy Beam Transport (LEBT) line in a modern ion accelerator typically connects an Ion Source (IS) to a Radio-Frequency Quadrupole (RFQ) (e.g. [1]). Its primary purpose is to properly match the beam's Twiss functions to those needed for injection into the RFQ. It may include a chopping system to introduce the appropriate time structure to the beam or shorten the rise and fall times of the beam pulse produced by the IS. The LEBT may also incorporate beam diagnostics, from a single beam current measuring device to multiple and more complex instruments.

As the beam propagates through a LEBT, residual gas molecules are ionized by the beam ions, and secondary particles can be trapped in the beam potential, thus neutralizing the beam's space charge (see for example [2]). This phenomenon affects the beam dynamics, in particular for the transport of low-energy, high-current beams. In fact, a typical design for a light-ion, high-intensity LEBT includes 1-3 solenoidal lenses for focusing [3] and relies on transport with nearly complete beam space charge neutralization over the entire length of the LEBT.

Often, a LEBT either operates with a pulsed ion source or creates pulses from an initially DC beam. Because the ionization process is not instantaneous, the front of the beam pulse is not neutralized as it propagates through the LEBT and may have Twiss parameters significantly different from their steady state values (e.g. [4]). Thus, for long-pulse operation, when the accelerator optics is tuned for neutralized transport in the LEBT, the space charge at the beginning of the pulse may result in increased losses in the LEBT and the following beam line.

Remedies include working at an increased residual gas pressure to speed up the neutralization process and moving the chopping system as close as possible to the RFQ in order to decrease the distance that the beam travels with full space charge and low energy. As an alternative solution, in this paper, we discuss the possibility and rationality of imposing un-neutralized beam transport dynamics in a long portion of the LEBT. In this alternative scheme, the ion source works in DC (or long pulse) mode and the beam propagates through the first, 'high pressure' part of the LEBT being neutralized, but neutralization is stopped by means of an electrical potential barrier and ion clearing device. Hence, in the ideal case with no neutralizing ions in the downstream part of the LEBT, the beam envelope is time independent. In this paper, we discuss conditions for such scheme not to introduce additional emittance growth and present its practical realization at the LEBT of the Proton Improvement Plan II Injector Test accelerator (PIP2IT) [5]. Consequently, further discussion and numerical estimations are made for the PIP2IT case: H- axially symmetric beam focused with solenoidal lenses, 5 mA, 30 keV, and 0.2 mm·mrad normalized emittance (rms).

Section II discusses effects of space charge dominated transport dynamics, which lead to the proposed transport scheme in Section III. Section III also includes simulation results that support the reasoning and conclusions brought forward. Section IV describes the PIP2IT LEBT, where measurements were made. Section V shows measurements substantiating the claim that the transport scheme with an un-neutralized section has been successfully implemented. The results and applicability of the scheme for cases where the beam has higher perveance are discussed in Section VI, followed by a summary.

## II SPACE CHARGE EFFECTS

Independently from the choice of the focusing elements technology, applicability of the proposed scheme depends on the beam's own space charge fields.

The overall strength of the space charge is typically characterized by the beam perveance defined as

$$P_b = \frac{I_b}{U_{IS}^{3/2}} \quad (1)$$

where $I_b$ is the beam current in Amps and $U_{IS}$, the ion source bias voltage (i.e. the beam's kinetic energy divided by the ions electrical charge $e$) in Volts. If the perveance exceeds a certain limit, an un-neutralized beam simply cannot be transported with lumped focusing.

Then, even if the perveance is not a limiting factor in that sense, the shape of the particles distribution may lead to an unacceptable increase of the beam emittance due to space charge forces.

[#]lprost@fnal.gov

## II.1 Linear space charge effect

Let us consider the space-charge dominated transport of a non-relativistic, round, completely un-neutralized H- beam with uniform transverse distribution of the charge density, i.e. neglecting the beam emittance and change of the longitudinal velocity across the beam. In this case, the maximum length, $L_m$, that the beam can propagate between two thin focusing elements while remaining within a maximum radius $r_b$ (in the middle of each lens) can be expressed (using, for example, formulae from Ref. [6]) as

$$\frac{L_m}{r_b} = 2.16 \cdot \sqrt{\frac{1}{2P_{Generalized}}} \quad (2)$$

where $P_{Generalized}$ is the "generalized" perveance, the unit-less equivalent to $P_b$, such that

$$P_{Generalized} = P_b \cdot \left(4\pi\varepsilon_0 \cdot \sqrt{\frac{2e}{M_i}}\right)^{-1/2} \quad (3)$$

with $e$ is the elementary charge in Coulombs, $M_i$ the ion mass in kg and $\varepsilon_0$ the permittivity of vacuum in SI units.

To limit spherical aberrations and other non-linearities of the solenoidal field, the inner radius of a typical magnetic lens is at least twice the anticipated beam radius, while typically its length is roughly equal to its bore diameter. One can argue that the maximum allowable perveance with lumped focusing corresponds to the case when the minimum practical physical distance between lenses exceeds their length by only a factor of 2-3 (with a factor of '1' meaning that the lenses would be touching). Then, taking a factor of 3 for the lenses' distance-to-length ratio implies

$$\frac{L_{m\_sc}}{r_b} \cong 12 \quad (4)$$

and the maximum allowable perveance is

$$P_{bm} \cong \left(\frac{r_b}{L_{m\_sc}}\right)^2 \times 3.59 \frac{\mu A}{V^{3/2}} = 0.025 \frac{\mu A}{V^{3/2}} \quad (5)$$

The nominal perveance at the PIP2IT LEBT for the 5 mA, 30 kV H- beam is $0.95\times10^{-3}$ μA/V$^{3/2}$, significantly lower than the estimation from Eq. (5). Therefore, in this simple model, un-neutralized transport is not excluded.

It is instructive to compare the estimate above with the opposite limit case of the emittance-dominated transport. The maximum distance between thin lenses where the beam rms size remains below $\sigma_{r0}$ can be estimated from formula (3.162) in Ref. [7] as:

$$L_{m\_em} = \frac{\sigma_{r0}^2}{\varepsilon} \quad (6)$$

where $\varepsilon$ is the beam rms emittance (un-normalized). Then, comparing the maximum distance that the beam can propagate between thin lenses for the space charge-dominated case and the emittance-dominated case, with $2\sigma_{r0} = r_b$ (by definition of the rms beam size of a uniform-density beam) and $P_b = 0.95\times10^{-3}$ μA/V$^{3/2}$, one finds that as long as $\sigma_{r0}/\varepsilon \gg 130$, the limitation to that distance in the un-neutralized case is dictated by space charge and not the beam emittance. Using $\sigma_{r0} \sim 7.5$ mm and $\varepsilon \sim 25$ μm, typical values for the PIP2IT LEBT, one gets $\sigma_{r0}/\varepsilon = 300$, and can conclude that the limitation on implementing lumped focusing is more restrictive due to space charge defocusing than the value of the emittance.

## II.2 Non-linear space charge effect

While the linear space charge effect is one factor pertaining to the applicability of space charge dominated transport, another limitation is the possibility of large emittance growth due to non-linear space charge forces. As an illustration, Figure 1 shows the emittance growth as a function of the beam current for a round beam with initial Gaussian distributions in all transverse planes (referred below as double-Gaussian) transported between two thin focusing elements within a fixed aperture. The initial convergence angle is adjusted for each beam current to keep the rms beam size at the end equal to its initial value.

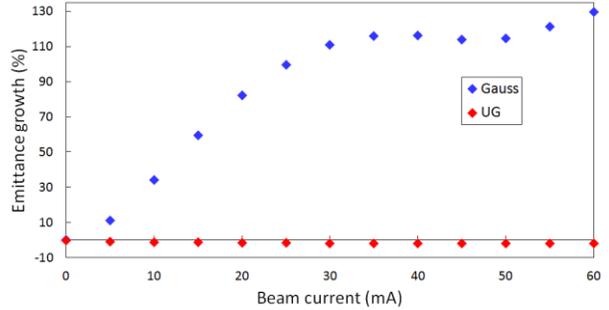

Figure 1: Emittance growth as a function of the H- beam current simulated with TraceWin [8]. The beam energy is 30 keV. Blue - Double-Gaussian initial distribution; Red - UG initial distribution. The drift length is 251 mm, the initial rms beam size is 7.5 mm, and the initial angle is adjusted for each current to provide the final rms size equal to the initial's. Initial beam emittance is 25 μm (rms, un-normalized) for both distributions. There is no beam loss.

Obviously, such emittance growth does not exist in the case of a constant space charge density beam propagating through a drift space like for a K-V distribution. However, it is not clear how such distribution can be realized in a LEBT.

A much more realistic particle distribution is one which is uniform in space and Gaussian in velocities (i.e. [9]):

$$f(x, x', y, y') = \frac{\hat{n} r_b^2}{\pi \varepsilon^2} \Theta\left[1 - \left(\frac{x^2}{r_b^2} + \frac{y^2}{r_b^2}\right)\right] \\ \times \exp\left[-\left(\frac{r_b x' - r_b' x}{\varepsilon}\right)^2 - \left(\frac{r_b y' - r_b' y}{\varepsilon}\right)^2\right]. \quad (7)$$

Here, $\Theta$ is the Heaviside unit-step function and $r_b'$ the beam initial angle (edge). The density $\hat{n} = \int f dx' dy' = const$ of the initial distribution is uniform within a circle of radius $r_b$. In Ref. [10], the authors simulated the propagation of such distributions through a channel with constant focusing and found no emittance growth. Note that in Ref. [10], this distribution is termed "semi-Gaussian", but in this paper, we call it Uniform-Gaussian, or UG. Similarly, the simpler model of propagation in a drift space between 2 thin lenses with initial UG particles distribution does not reveal an emittance growth at any simulated beam currents (Fig. 1). Note that the maximum current (60 mA) is determined by the linear space charge limit, which corresponds to the case where the final beam size cannot be made equal to the initial beam size for any Twiss function $\alpha$, for a fixed distance between the lenses.

These examples allow to presume that in a real LEBT the space charge-induced emittance growth can be suppressed if the initial distribution is close to UG. Some additional considerations that brought us to this idea can be found in Ref. [11].

## III SCHEME OF LEBT WITH UN-NEUTRALIZED SECTION

### III.1 A hybrid transport scheme

As mentioned previously, neutralized transport with solenoids of low-energy, high-intensity, light-ion beams is the norm. On the other hand, as indicated in Section II, if the perveance of the beam is modest, space-charge dominated transport is possible and the emittance growth might be insignificant if the beam current density distribution where neutralization is interrupted is close to UG. In practice, plasma-based ion sources release a significant amount of gas into the vacuum chamber, so that at least partial neutralization of the extracted beam is inevitable. Note that partial neutralization might be the worst scenario for beam transport since in this case the neutralizing ions are unevenly distributed across the beam and create highly non-linear electric fields, which dilute the emittance.

Thus, ideally, the LEBT should accept a neutralized beam and transport it some distance away from the ion source at which location neutralization is removed. Then, the beam propagates and is matched to the RFQ fully un-neutralized.

### III.2 Image formation with an axially symmetrical solenoidal lens

In the proposed transport scheme, a fully un-neutralized beam is assumed to propagate over a relatively long section of the LEBT. However, examples in Section II.2 hint that, unless the initial beam current density distribution is uniform, non-linear space charge forces will lead to a significant emittance growth.

Let's consider a beam focused by an axially symmetrical solenoidal lens. For a completely neutralized beam transport with linear optics, the lens replicates the initial current density distribution in the image plane. Consequently, if the beam current density at the ion source exit is uniform, the beam will have a uniform current density distribution at some distance downstream of the lens, but magnified with respect to the one that was generated (Fig. 2). Position of the image corresponds to the betatron phase advance of π. Note that in the middle the distribution can be very far uniform. For example, if at some location, the beam size is increased by a large factor due to thermal velocities, particle's position is determined mainly by its initial transverse velocity, and the distribution is closer to Gaussian (Fig. 2b).

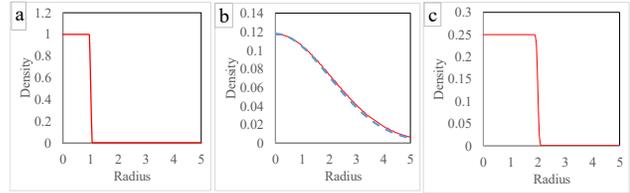

Figure 2. Illustration of the evolution of a beam with initial uniform current density distribution (a) as it expands due to thermal velocities, propagates through a thin focusing lens (b), and returns to a uniform density distribution in the image plane of that lens (c). The plots show analytical calculations of the current density distributions in MathCad. Density and radius are normalized by the initial density in the center of the beam and the initial beam radius. The distance between the lens and the image plane is twice the distance between the emitter plane and the lens. The dash blue line in (b) shows a Gaussian fit.

Note that the scale of the distance over which the emittance growth due to space charge occurs is defined by the beam's plasma length, the latter being proportional to the rms beam size and inversely proportional to the beam perveance in the paraxial approximation with continuous focusing [10]. Therefore, it is preferable to have a large beam radius at the transition to un-neutralized transport, i.e. a magnification > 1.

Figure 3 illustrates the same concept as Figure 2 with results of PIC-like simulations of a beam transported through one solenoid with a realistic longitudinal magnetic field profile. The initial particle distribution is UG. The beam current density distribution approaches Gaussian as it propagates but is uniform again in the image plane.

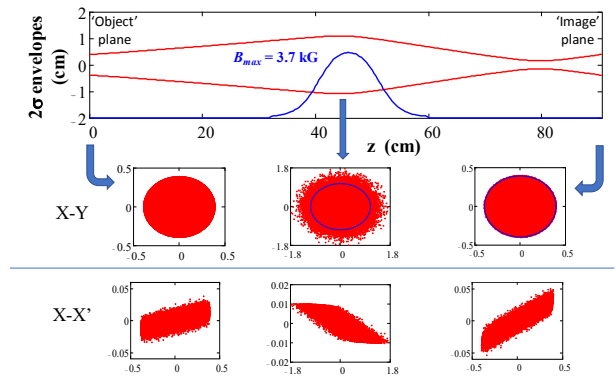

Figure 3. Numerical simulation with VACO [12] of the propagation of a beam with the initially uniform spatial density distribution and Gaussian velocity distribution through a magnetic lens. For the phase space portraits, units are centimeters for X and Y, and radians for X'.

### III.3 Conceptual realization of the transport scheme with an un-neutralized section

Based on the considerations above, the following LEBT scheme with an un-neutralized section is proposed:
- An ion source is optimized to generate a uniform spatial density distribution;

- The tails of the distribution, which are usually not uniformly populated, are scraped off after the beam is let expanding as it propagates toward the first solenoid;
- The beam transport from the ion source through a first solenoid is as close as possible to being completely neutralized;
- The beam size near the image plane of the first solenoid is increased to the limit imposed by aperture limitations of critical elements downstream, e.g. chopping system;
- Near that image plane, neutralization is interrupted;
- The length of the remaining portion of the LEBT is minimized, and concentration of thermal ions there is kept as low as possible by combining ion clearing and good vacuum.

A conceptual realization of the scheme based on the PIP2IT lattice configuration is shown on Figure 4.

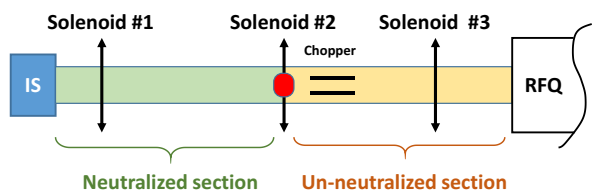

Figure 4: Transport scheme schematic.

The transition to un-neutralized transport is achieved by the combination of a potential barrier (red oval on Fig. 4) that confines thermal ions upstream and a clearing electric field created by a negative DC offset on one of the chopper plates that sweeps them out of the beam path downstream. In addition, a low vacuum pressure maintained between the potential barrier and the RFQ limits the rate at which neutralizing particles are created.

It should be noted that realization of the scheme places restrictions on the design of the beam line. Specifically, a 3–solenoid scheme is preferable: the first solenoid allows adjusting, for a specific regime of the ion source, the position of the image plane to the location of the transition from neutralized to un-neutralized transport; the other 2 solenoids match the beam to the RFQ.

### III.4 Simulations

The scheme depicted in Figure 4 was simulated with TraceWin keeping in mind its practical realization at PIP2IT, where the chopping system installed between the second and third solenoids determines the allowable beam size in this region.

To simplify the simulations, we assume that the phase advance from the ion source emitting surface to the position downstream of the ion source where neutralization is complete is small and can be neglected. Consequently, the simulations begin with ions at full energy and complete neutralization. While in the real beam line this assumption is likely incorrect, the additional phase advance would only result in a shift of the optimum position of the image plane, not affecting the main conclusions. In addition, in this model, the scraping that takes place at the exit of the ion source vacuum chamber in the actual machine (section IV.1) is not implemented. Hence, the input emittance is chosen to be lower than actually measured (Case A in Table 1).

Results of numerical simulations are presented in Figure 5 for two cases, which differ by the choice of the initial current density distribution: UG in one case, and double-Gaussian in the other. These simulations use 3D fields obtained from Microwave Studio [13] using the Opera [14] model of the solenoids.

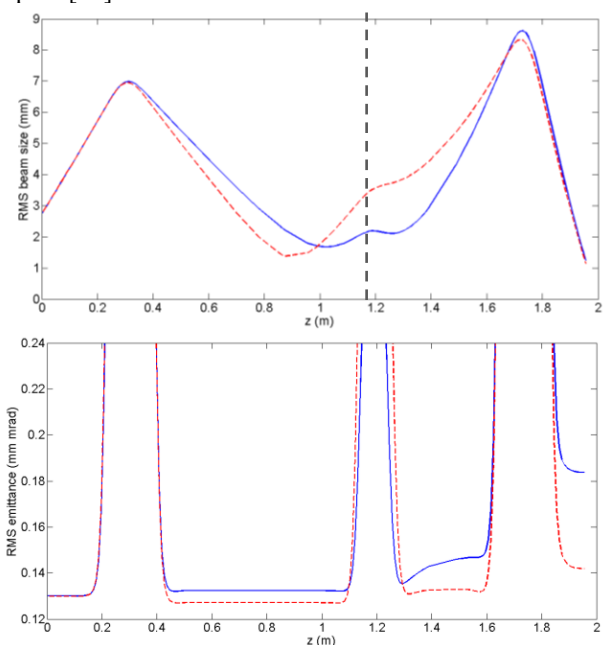

Figure 5: Top - Beam envelopes simulated with TraceWin starting with a beam distribution with uniform current density (dashed, red) and a Gaussian current density distribution (solid, blue). The dashed vertical line indicates the transition from neutralized to un-neutralized transport. Bottom - Corresponding emittance evolution. Apparent jumps of the emittance inside the solenoids are related to the specific way of emittance calculation by TraceWin.

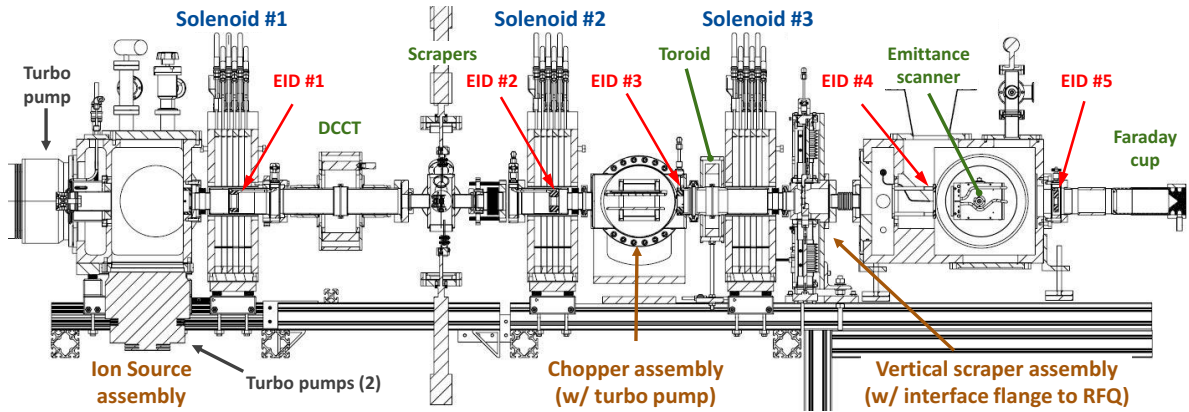

Figure 6: PIP2IT LEBT beam line (side view).

Note that the focusing strengths of the solenoids are adjusted such that the Twiss parameters at the end of the beam line are nearly identical, while also maintaining the overall shape of the envelope as it propagates. Also, the slight decrease of the rms emittance seen as the beam propagates through Solenoid #1 in the UG case might be explained similarly to the reasoning from Ref [10].

Thus, Figure 5 shows that when the initial current density was chosen to be uniform, there is little emittance growth (<10%), while the emittance grew by ~40% when the initial current density distribution was chosen to be Gaussian.

## IV REALIZATION

In the context of the Proton Improvement Plan II (PIP-II) program at Fermilab [15], a test accelerator PIP-II Injector Test, or PIP2IT (previously known as PXIE), is being constructed to address various technical risks associated with the design of the low energy end of the PIP-II SRF linac, including the LEBT transport scheme with an un-neutralized section.

### IV.1 Beam line layout

A layout of the PIP2IT beam line, as configured for the measurements reported below, is shown on Figure 6. In addition, Figure 7 presents a simplified schematic view of the same with dimensional information (e.g. distance between lenses, aperture limitations). The beam is generated by a commercial H- Volume-Cusp Ion Source [16], with a modulator circuitry added to its extraction electrode. The beam line consists of 3 solenoids; a set of 4 transverse, radiation-cooled scrapers (installed as temporary diagnostics between solenoids #1 and #2); a chopping system; Electrically Isolated Diaphragms (EID) (water-cooled, except for EID #4); a vertical scraper assembly (an electrically-isolated, water-cooled, movable electrode with 3 apertures of different sizes), and current diagnostics [17]. The chopper assembly consists of a 1000 l/s turbo pump, an electrostatic kicker and an EID. The kicker has two plates: one is grounded (and electrically-isolated for current measurements) and also serves as the beam absorber; the second is biased to -5 kV to deflect the beam to the absorber, and brought towards ground to pass the beam. Details about most of the components can be found in [18]. EIDs #1 and #2 are typically biased to +50 V to prevent background ions from moving from one section of the LEBT to another, while positive ions are cleared in the last ~1 m of the beam line before the RFQ by applying -300V DC voltage to the kicker plate. An Allison-type emittance scanner can be installed either at the end of the beam line as shown in Figure 6 or in the Ion Source assembly's vacuum chamber.

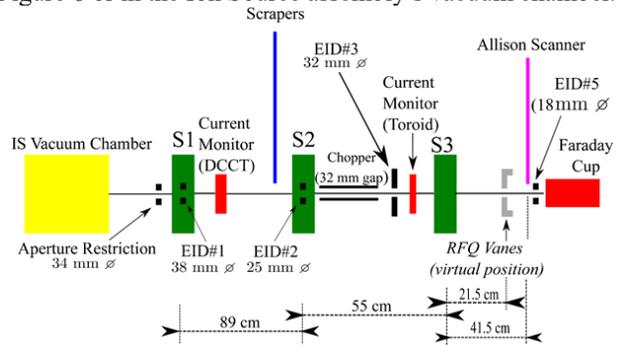

Figure 7: Beam line schematic. 3" beam pipe (OD) unless otherwise noted; Faraday cup aperture is Ø 67 mm.

It should be noted that an aperture restriction at the exit of the ion source vacuum chamber significantly collimates the beam. For the nominal IS settings used to obtain 5 mA at the DCCT, we measured that ~20% of the beam is scraped off. Simulations of the beam transmission through the 1st solenoid made with TRACK [19] agree to within 5%. In addition, they indicate that the beam emittance decreases noticeably, by as much as 35% for this particular case [20].

### IV.2 Vacuum profile along the beam line

The conceptual design of the transport scheme assumes a specific vacuum distribution along the beam line: relatively high pressure at the beginning to favour neutralization, and good vacuum in the section where un-neutralized transport is desired. With the ion source employed at PIP2IT, in the extraction region, the vacuum pressure is at least 2 orders of magnitude higher than in the vicinity of the chopper, where it is of the order of $1\times10^{-7}$ Torr. A simulation using the Molflow code [21] (Fig. 8) shows that the pressure in the ion source region drops from $1.3\times10^{-2}$ Torr at the plasma surface to $1.1\times10^{-5}$ Torr at the downstream end of the vacuum chamber. The vacuum pressure measured with an ion gauge located relatively far from the beam path is in overall good agreement with this simulation, routinely reading $4\text{-}8\times10^{-6}$ Torr. From the exit of the ion source vacuum chamber to the chopper assembly, where the vacuum with DC beam deflected onto the absorber is typically $\sim10^{-7}$ Torr, a linear decrease of the vacuum pressure can be assumed. Then, it remains nearly constant with the last pumping station located in the chamber that houses the emittance scanner.

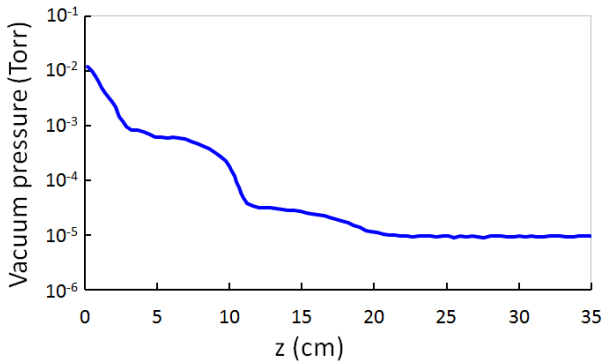

Figure 8: Molflow simulation of the axial pressure longitudinal distribution in the ion source region of the PIP2IT LEBT. $z = 0$ indicates the plasma emitting surface. An $H_2$ gas flow of 15 sccm was used.

### IV.3 Neutralization pattern

In the simplest model, where neutralizing ions are confined longitudinally, the space-charge compensation builds up linearly until reaching an equilibrium determined by the balance between the radial loss of the neutralizing ions and their production, the latter being driven by the local vacuum pressure. The time to reach such equilibrium (a.k.a. neutralization time) is determined only by the residual pressure and given by

$$\tau_{comp} = (n_{gas}\,\sigma_i\,v_p)^{-1}, \qquad (7)$$

where $n_{gas}$ is the gas density, $\sigma_i = 1.5\times10^{-16}$ cm$^2$ [22] the ionization cross section of the $H_2$ gas, and $v_p$ the velocity of the $H^-$ ions. For the PIP2IT LEBT vacuum profile discussed in Section IV.2, the neutralization time given by Eq. (7) varies from tens of microseconds in the ion source vacuum chamber to milliseconds downstream of the chopper.

In a more realistic model, neutralizing ions can travel along the beam line and their confinement within a certain location is dictated by the potential distribution along the LEBT. At PIP2IT, this distribution is the superposition of the beam potential and electric fields of EIDs. Measurements of the leakage current to the chopper plates while varying the biasing potential of the EIDs supported that model [20]. Thus, with the biasing scheme and vacuum profile presented previously, the neutralization pattern should be reasonably close to the ideal case of Figure 4 where a step function is assumed.

## V PROOF OF PRINCIPLE

### V.1 Beam distribution at the ion source

The Ion Source is commercially available and not necessarily optimized to deliver a beam with a uniform current density distribution. On the other hand, it seems natural to expect the beam current density distribution coming out of the ion source emitter to be closer to uniform rather than Gaussian. At the same time, the beam formation out of a plasma in a near thermal equilibrium must result in a Gaussian velocity distribution.

Information about the current density distribution can be extracted from phase portraits recorded with the emittance scanner installed near the ion source. Assuming the beam drifts in free space with no space charge, the phase space distribution can be propagated back toward the ion source with a simple linear coordinate transformation for each cell of the recorded distribution. Figure 9 shows an example of such back-propagation to the location of the ion source ground electrode, with cells distributed over 40 bins along the position coordinate.

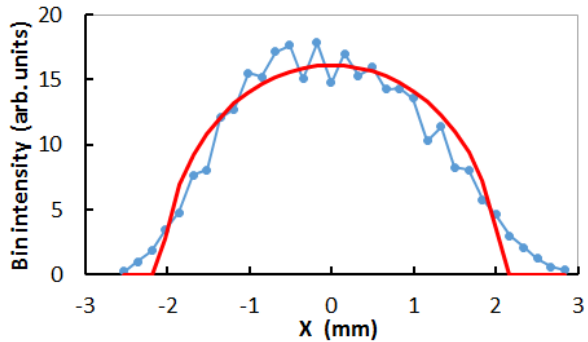

Figure 9: Comparison of a 1D current density distribution from a back propagated phase portrait (blue) and a uniform distribution fit (red).

The distribution displays features consistent with both a radially uniform distribution (1D projection, also plotted on Fig. 9) in the core and Gaussian-like tails. The collimation that takes place at the exit of the ion source vacuum chamber should remove the tails, producing at the image plane of Solenoid #1 a beam with a nearly uniform spatial distribution favourable to the transport scheme described here.

### V.2 Choice of the pulse structure

The profile measurements described below were performed at the end of a 1.5 ms pulse formed by the IS modulator. As such, the modulator pulse is assumed to be long enough to allow reaching a steady state upstream of EID #2. Emittance measurements at the end of the LEBT were carried out with a 50 μs pulse chopped out at the end of the modulator pulse so that the chopped pulse is much shorter than the typical neutralization time downstream.

### V.3 Profile measurements

Feasibility of the proposed scheme is based on the assumption that the beam density profile at the transition between neutralized and un-neutralized regions can be tuned to be nearly uniform, while at an arbitrary focusing it would be between uniform and Gaussian. Accurate measurement of the profile at this location, which occurs inside Solenoid #2, is difficult. Instead, the transition between profile shapes is observed slightly upstream of EID #2 with scrapers (see Figs. 6 & 7 for location). Figure 10 shows such profiles (1D) for two cases with different Solenoid #1 currents, $I_{Sol\ \#1}$, but the same IS settings. Best fits (in the least squares sense) from analytical expressions of the profiles derived by employing either a Gaussian or uniform current density distribution are displayed as well. For $I_{Sol\ \#1}$ = 155 A, assuming a uniform current density distribution visually leads to a better fit than Gaussian, which is not the case for $I_{Sol\ \#1}$ = 140 A.

Quantitatively, for each profile and either distribution function, one can compute the Root-Mean-Square Error (RMSE) of the fits:

$$\text{RMSE}_k = \sqrt{\frac{\sum_{i=1}^{n}(y_{Data_i} - y_{Fit_i})^2}{n}} \quad (8)$$

where $n$ is the number of data points and $k$ indicates, which of the 2 profiles is being considered. For $I_{Sol\ \#1}$ = 155 A, RMSE$_{155}$ is 0.024 and 0.081 mm assuming a uniform and Gaussian current density distribution, respectively, which corroborates the conclusion made from merely looking at the graphs. Similarly, for $I_{Sol\ \#1}$ = 140 A, RMSE$_{140}$ is 0.044 and 0.062 mm, respectively. While a uniform current density distribution remains a better model than Gaussian for this solution, both the fact that RMSE$_{140}$ > RMSE$_{155}$ for the uniform case, and RMSE$_{140}$ < RMSE$_{155}$ for the Gaussian case, indicate that the beam current density distribution for $I_{Sol\ \#1}$ = 140 A deviates more from uniformity than for $I_{Sol\ \#1}$ = 155 A. Thus, based on the concept described in the previous section, one may argue that if the transition from neutralized to un-neutralized transport would occur near the location of the scrapers, the beam emittance would stay nearly constant downstream for $I_{Sol\ \#1}$ = 155 A, while it would deteriorate for $I_{Sol\ \#1}$ = 140 A.

Simulations confirm that the current density distributions are very different at the location of the scrapers for the 2 values of $I_{Sol\ \#1}$ and are consistent with the data. This is illustrated on Figure 11, which compares the derivatives of the profiles presented in Figure 10 with the corresponding ones obtained from TraceWin. Note that for the simulations, the reconstruction of the beam profile was such that the bin size was equal to the step size in the measurements, which, explains why the profiles are not smooth.

Also, while the overall shapes and trends of the profiles agree well between the data and the simulations, quantitatively, the beam sizes do not. Simulation results consistently underestimate the beam radius $\sigma$ by ~26%. This is believed to be due to a combination of measurement errors and incomplete neutralization upstream of Solenoid #2.

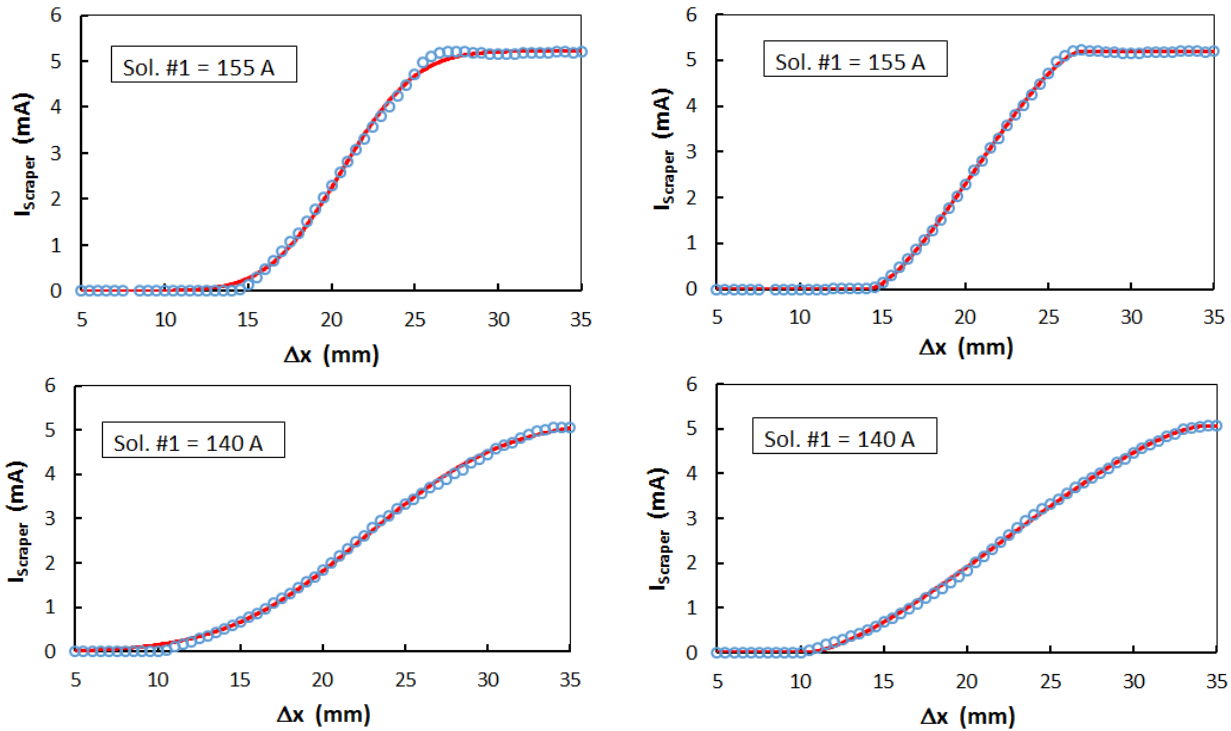

Figure 10: 1D Beam profiles recorded with the "right" scraper (horizontal displacement). Top row: For Solenoid #1 = 155 A; Bottom row: For Solenoid #1 = 140 A. The data points are the blue circles. In red are fits from an analytical formulation assuming that the beam current density distribution is either Gaussian (left column) or uniform (right column).

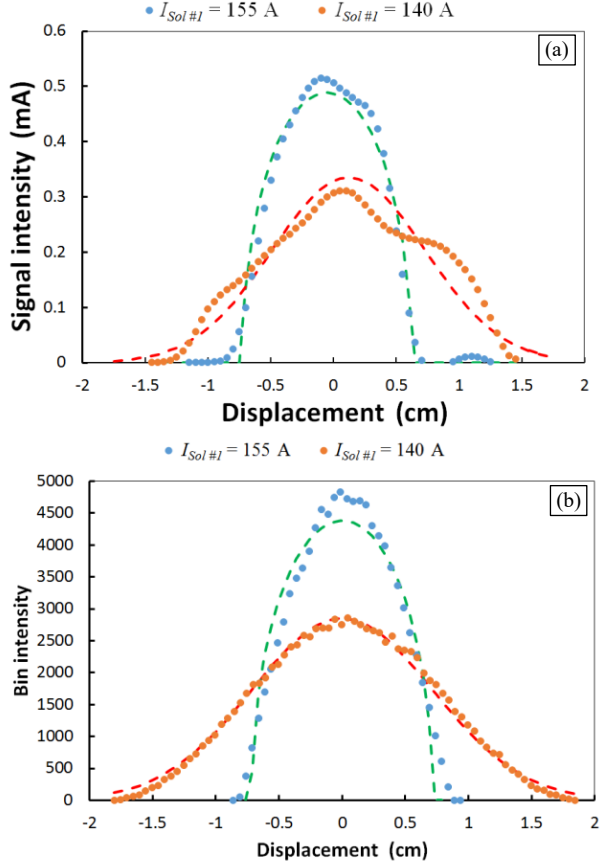

Figure 11: 1D beam profiles at the scrapers. a - Data; b - TraceWin simulations. Dashed curves are fits, assuming a uniform (green) or a Gaussian (red) distribution.

*V.4  Emittance measurements*

Low emittance beams were measured at the end of the PIP2IT beam line under various biasing and focusing configurations [23]. Table 1 shows the results of 3 particular phase space measurements of interest: one at the exit of the ion source (A) (*upstream* of the IS vacuum chamber aperture restriction where a non-negligible fraction of the beam is scraped off) and two downstream of Solenoid #3 (B & C). For all cases, the ion source was tuned identically and the EIDs biasing configuration was the same. The beam current as read by the DCCT is 5 mA. The data downstream is for a 50 μs chopped pulse, which, as mentioned previously, is much shorter than the neutralization time in that section of the beam line, hence a fair representation of an un-neutralized beam. As shown in the table, while focusing settings are significantly different, the measured Twiss parameters are nearly identical at the end of the beam line. Nevertheless, the measured emittances in all 3 cases are different.

Table 1: Phase space measurements results

|   | Sol. #1 [A] | Sol. #2 [A] | Sol. #3 [A] | $\varepsilon_n$ (rms) [μm] | $\alpha$ | $\beta$ [m] |
|---|---|---|---|---|---|---|
| A | – | – | – | 0.19 | -3.5 | 0.6 |
| B | 154 | 187 | 223.5 | 0.25 | -8.9 | 2.2 |
| C | 143 | 158 | 240 | 0.16 | -8.2 | 2.2 |

We explain the decrease of the emittance between A & C by the scraping that takes place at the exit of the IS vacuum chamber (section IV.1). At the same time, the fact that, for B, the emittance at the end of the LEBT is significantly larger than measured at the ion source exit (A) clearly shows that scraping alone does not necessarily lead to a beam with low emittance downstream. Conversely, space charge dominated transport does not necessary cause unacceptable emittance growth (measurement C, where the emittance at the end of the beam line is actually lower than at the exit of the ion source). Our interpretation is that measurement B corresponds to the case where the beam current density is significantly far from uniformity near EID #2, while it is close to being uniform for measurement C. Note that it is merely coincidental that the value of $I_{Sol\ \#1}$ for measurement C is nearly identical to the one that leads to a smaller RMSE at the scrapers for the Gaussian distribution than for the uniform distribution (Fig. 11, orange circles).

Also, spherical aberrations from the solenoidal lenses, which can significantly deteriorate the beam quality in some instances, have been deemed negligible. First, they were estimated analytically and then calculated from magnetic simulations and found to contribute to <5% to the emittance growth [24]. Second, the phase space at the exit of the LEBT is not distorted, even when the beam was intentionally made large in the last transport solenoid. Finally, the Solenoid #3 current is actually larger in measurement C (no emittance growth) than in measurement B; if spherical aberrations were responsible for the emittance growth observed in measurement B, one would expect measurement C to be even worse.

Therefore, we believe that we have reasonable evidence that the transport scheme with an un-neutralized section was realized and exhibit the properties enumerated in Section III.3, to within the uncertainties associated with a real accelerator.

## VI  DISCUSSION

With the experimental results indicating that the suggested transport scheme is working, it is useful to discuss two questions:
- What are the practical benefits and drawbacks of the scheme?
- Can this scheme be used for a higher beam perveance?

*VI.1  Scheme practicality*

The PIP2IT LEBT delivers a beam with the required parameters to the RFQ while preserving a low emittance. Avoiding the requirement of full neutralization allows maintaining a good vacuum in the RFQ ($2\times10^{-7}$ Torr with a full–intensity beam), which should improve the long-term reliability of the RFQ operation. Note that for the case of H- beams, a pressure artificially elevated for neutralization results in additional beam losses due to stripping.

Measurements of the beam accelerated by the RFQ showed that, with the nominal bias voltage of +50 V on

EIDs #1 and #2, beam parameters still vary in time to within 10% over a 0.5 ms chopped pulse. For a beam current of 5 mA, further optimization by increasing the bias of EIDs #1 and #2 to +100V and using a DC beam from the IS made the beam parameters constant during the pulse within measurement errors (<5%) [25]. One of benefits from having steady beam properties throughout the pulse is that all tuning can be performed with short pulses (in practice, 10 μs), combining a high degree of fidelity for predictions of the beam behaviour for long-pulses with a very safe mode of operation.

On the other hand, the proposed transport scheme does add complexity to the LEBT design. Probably, the most significant requirement is to create a UG distribution. In the specific case of the commercially available PIP2IT ion source, a noticeable portion of the beam (tens of %; ~20% in our specific case) needs to be scraped off to eliminate the tail particles deviating from the ideal UG distribution. Also, the scheme requires an electrode that stops the neutralizing ions (EID #2 in Fig. 6) and an ion clearing voltage at the chopper plate while passing the beam through. Finally, making the locations of the image plane and stopping electrode coincide is the third requirement in addition to matching of the two Twiss parameters at the RFQ entrance. Flexibility of matching for any IS tune implies that three solenoids are needed. Note that in the PIP2IT case, where a 3-solenoid scheme was chosen initially to simplify transport through a bend installed later between Solenoid #1 and #2, the placements and strengths of the solenoids do not create significant tuning restrictions. In part, fully neutralized transport without ion clearing has been demonstrated with the same output emittance as in case C of Table 1.

### VI.2 Applicability to cases with higher beam perveance

The transport scheme with an un-neutralized section was realized in the PIP2IT LEBT at the beam current of 5 mA (perveance of $9.5 \times 10^{-4}$), which is at the lower end of the overall range of LEBTs [3]. However, in simulations we did not find restrictions for applying it to higher perveance cases solely based on an eventual emittance growth due to non-linear space charge fields. In the example of a UG beam propagating in the drift space (Fig. 1), the emittance growth remains negligible over the entire range where the transport is allowed by the linear space charge effects (i.e. containment of the beam envelope within a given aperture).

A case closer to a real LEBT is illustrated in Figure 12 below showing results of a simulation of the scheme using VACO with partially un-neutralized transport for a beam perveance of $4.1 \times 10^{-3}$ (100 keV, 140 mA), more than 4 times larger than at the PIP2IT LEBT. At the LEBT exit, the emittance increases by less than 5%. Therefore, an artificial pressure rise at the RFQ entrance and in-pulse variations can be avoided even at significantly higher perveance values if the additional efforts mentioned in Section VI.1 are deemed reasonable.

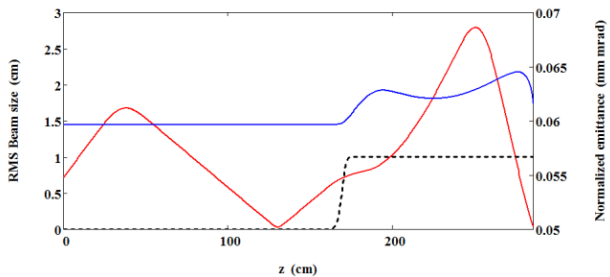

Figure 12: Simulation results from VACO of an example with a high perveance beam ($4.1 \times 10^{-3}$) – the initial particle distribution is UG. Red - RMS beam size; Dashed black: Neutralization profile (0 ≡ Neutralized, 1 ≡ Full beam current); Blue - Emittance evolution (normalized).

Note that all discussions in this paper assume that space charge–related emittance growth is dominant. If aberrations of focusing solenoids are large, as reported in Refs. [26, 27] for the same parameters as in Figure 12, the phase portrait can be dramatically distorted, and benefits of the UG distribution cannot be realized. Also, beside limitations due to optics imperfections, difficulties with preparing a UG distribution may increase with the absolute value of the beam current and power in or near the ion source. However, in the case of modest–power beams, we would expect the transport limit to be dictated by linear effects, therefore, the scheme presented in this paper might be applicable to most LEBTs.

## VII SUMMARY

An atypical transport scheme for low-energy, high-intensity, light-ion beams was devised and implemented. Its originality resides in the fact that it incorporates a relatively long section where the beam is un-neutralized, while it exits the ion source highly neutralized. Physical arguments and numerical simulations demonstrate that if the transition from neutralized transport to un-neutralized transport occurs at a location where the beam current density is nearly uniform, emittance growth downstream is greatly reduced.

The scheme with an un-neutralized section was realized at the PIP2IT LEBT. Phase-space and profile measurements of the beam along the beam line support features and behaviours obtained in the corresponding simulations. In particular, a low-emittance beam was successfully produced at the end of the LEBT.

Ultimately, the proposed transport scheme allows combining low emittance growth, low pressure at the RFQ entrance, and constant beam properties throughout the pulse. While it is demonstrated for a comparatively low-perveance case, no theoretical limitations for applying it to significantly more intense beams, with perveance limited by the linear space-charge effect, are found.

## VIII ACKNOWLEDGEMENT

Authors recognize important contributions from B. Hanna, R. D'Arcy, V. Scarpine, and C. Wiesner in various experimental activities and data analyses at PIP2IT, and are grateful to the Accelerator Division